\documentstyle[11pt]{article}\textheight 230mm\textwidth 150mm
            \newcommand{\be}{\begin{eqnarray}}
            \newcommand{\ee}{\end{eqnarray}}
            \newcommand{\eel}[1]{\label{#1}\end{eqnarray}}
\newcommand{\e}[1]{\label{e:#1}\end{eqnarray}}
     \newcommand{\eg}{{\em e.g.\ }}
            \newcommand{\ie}{{\em i.e.\ }}
            \newcommand{\ga}{{\gamma}}
 
            \newcommand{\la}{{\lambda}}
            
            \newcommand{\del}{{\delta}}

           \newcommand{\ra}{{\rightarrow}}
 
            \newcommand{\lra}{{\leftrightarrow}}

            \newcommand{\beq}{\begin{quote}}
            \newcommand{\eq}{\end{quote}}
            
            \newcommand{\al}{\alpha}
            \newcommand{\ben}{\begin{enumerate}}
            \newcommand{\een}{\end{enumerate}}
            \newcommand{\bit}{\begin{itemize}}
            \newcommand{\ei}{\end{itemize}}
    	\newcommand{\nn}{\nonumber}
            \newcommand{\r}[1]{(\ref{e:#1})}
            \newcommand{\edfl}[1]{\label{#1}\end{df}}

\newcommand{\vb}{{\cal h}}
\newcommand{\hb}{{\cal i}}

\def\d{\partial}
\def\cC{{\cal C}}
  \def\bcC{\bar{{\cal C}}}
  \def\cP{{\cal P}}
 \def\cH{{\cal H}}
  \def\cL{{\cal L}}
  \def\bcP{\bar{{\cal P}}}
  \def\half{{1 \over 2}}
  \def\lm{\lambda}

\begin{document}
\begin{titlepage}
\noindent
G\"{o}teborg ITP 97-11\\
June 1997\\
hep-th/9706044\\

\vspace*{5 mm}
\vspace*{35mm}
\begin{center}{\LARGE\bf Generalized BRST quantization\\
and massive vector fields} \end{center} \vspace*{3 mm} \begin{center} \vspace*{3 mm}

\begin{center}Robert
Marnelius\footnote{E-mail: tferm@fy.chalmers.se} and Ikuo S.
Sogami\footnote{Address:  Department of Physics, Kyoto Sangyo University, Kyoto 603, Japan
\\ \ \ E-mail: sogami@cc.kyoto-su.ac.jp}\\ \vspace*{7 mm} {\sl
Institute of Theoretical Physics\\ Chalmers University of Technology\\
G\"{o}teborg University\\
S-412 96  G\"{o}teborg, Sweden}\end{center}
\vspace*{25 mm}
\begin{abstract}
A previously  proposed generalized BRST quantization on inner product spaces for
second class constraints is further developed through applications. This BRST method
involves a conserved generalized BRST charge $Q$ which is not nilpotent $Q^2\neq0$ but
which satisfies $Q=\del+\del^{\dag} $, $\del^2=0$, and by means of which physical
states are obtained from the projection $\del|ph\hb=\del^{\dag}|ph\hb=0$. 
A simple model is analyzed in detail from which some basic properties 
and necessary ingredients are extracted.
The method is then applied to a massive vector field. 
An effective theory is derived which is close to the one of the St\"uckelberg model.
However, since the scalar field here is introduced 
in order to have inner product solutions,
a massive Yang-Mills theory with polynomial interaction 
terms might be possible to construct. 
\end{abstract}\end{center}\end{titlepage}

\setcounter{page}{1}
\setcounter{equation}{0}
\section{Introduction.}
BRST quantization in the BFV formulation \cite{BFV} is a  consistent procedure
to quantize general gauge theories \ie systems with first class
constraints in Dirac's constraint classification. Now there are many interesting
models which naturally are formulated as systems with second class constraints.
Massive vector fields and massive Yang-Mills are here important examples. In this
paper we develop a generalized BRST quantization for systems of the latter type and
at the end we treat the massive vector field  as an example. The method raises some
hopes that massive Yang-Mills may be quantized in a consistent fashion without a
physical scalar boson.

Quantization of theories with second class constraints are usually avoided  although
such theories are possible to quantize. There are in fact three main roads to their
quantization: 1) One may eliminate the constraints after which one obtains a regular
quantum theory which, however, usually is no longer manifestly Lorentz invariant. A
consistent renormalization procedure for such field theories seems therefore
extremely difficult to set up.  2) One may formulate a modified theory which in a
particular sense is equivalent to the original theory by turning the second class
constraints into first class ones  by means of auxiliary fields. ( A general method
is given in \cite{BF}.) This modified theory may then be quantized as an ordinary
gauge theory.
 Notice, however, that the modified theory can never be exactly
equivalent to the original theory. It may \eg  have different
global symmetries. Massive Yang-Mills  may be turned into a gauge theory by means of
 a so called St\"uckelberg transformation
\cite{Stuck}, which introduces auxiliary scalar fields.
 However, the resulting Lagrangian is nonlinear in these St\"uckelberg
fields \cite{KGo} and is therefore nonrenormalizable \cite{ABB}.  3) The third
possibility is to quantize the second class theory as a projection from an original
state space spanned by the original variables:  i) One may use the path integral
formulation of Senjanovic \cite{PSe} in which simply the measure
$\sqrt{\det \{\phi_a,\phi_b\}}\prod_a \del(\phi_a)$ is inserted where $\phi_a=0$ are
the second class constraints.  There are also
more elaborate BRST approaches which are designed to avoid singular measures: ii) One
approach involves two anticommuting nilpotent charges, $Q^a,\;a=1,2$, which are used
to project out physical states according to $Q^a|ph\hb=0$ \cite{BTL1,BTL2}. The number
of ghost fields in this approach is half the number of second class constraints. This
method is applied to massive Yang-Mills in \cite{BTL1}. iii) Another approach
was proposed in \cite{RMa,Bigrad} and will be investigated  in this paper.

In \cite{RMa} one of us proposed the following generalized BRST formulation: If the
constraints satisfy an algebra of the general form
\be
&&[\phi_a, \phi_b]=iU_{ab}^{\;\;c}\phi_c+if_{ab},
\e{1}
where $f_{ab}$ does not vanish on the constraint surface $\phi_a=0$, then we have a
theory with second class constraints. A particular example is a gauge theory with
anomalies in which case $f_{ab}$ is the anomaly ($f_{ab}\propto\hbar$). The BRST charge is
then constructed as if $f_{ab}$ is zero using the BFV algorithm \cite{BFV}. The number
of ghost fields is then always equal to the number of constraints. As a consequence
the resulting BRST charge is not nilpotent. We have
\be
&&Q^2=\half i f_{ab}C^aC^b +\cdots,
\e{2} 
where $C^a$ are the ghost fields. If $Q$ is conserved one may perform the
projection
\be
&&Q|ph\hb=0.
\e{3}
However, this is not
a sufficient condition for a consistent theory as was shown in \cite{Oj} where
massive Yang-Mills was considered. In \cite{Bigrad} an additional condition was
proposed. This condition was introduced in order to have solutions which belong to
inner product spaces. The additional requirements were that  
the BRST charge must be possible to decompose as follows 
\be
&&Q=\del+\del^{\dag},\;\;\;\del^2=0 
\e{4} 
and that \r{3}  should be  replaced by 
\be
&&\del|ph\hb=\del^{\dag}|ph\hb=0,
\e{5}
where $\del$ and $\del^{\dag}$ are independent and contain effectively half the
constraints in $Q$ each. In the ordinary case when $Q$ is nilpotent \r{3} and \r{5}
are equivalent. However, when $Q^2\neq0$  \r{5} is more restrictive than \r{3}, in
other words \r{5} implies \r{3} but not vice versa. By means of a simple example it
was demonstrated that the problem with \r{3} as pointed out in \cite{Oj} is avoided by
\r{5}.

The above generalized BRST procedure for second class constraints is a straight-forward
generalization of the quantization procedure for general gauge theories  as formulated in
\cite{Bigrad},\cite{Simple}--\cite{Time}.
With the
conditions \r{4} and \r{5} it has so far not been 
applied properly to any model. It has
been applied to a trivial case in \cite{Bigrad}, and if one neglects the zero modes
it was successfully applied to the free anomalous string in subcritical
dimensions in \cite{RMa,String}. It has also partly 
been applied to the massive superparticle in \cite{MS}.
Now the above generalized BRST quantization is very poorly developed.
We have \eg no simple algorithm which determines effective theories
with generalized BRST invariance. In order to develop appropriate techniques 
we perform a detailed study of a simple model with finite number of degrees of
freedom in section 2. This model contains some typical 
features which make the
results generalizable to more involved models. 
In section 3 the results are applied to the massive vector field.
We find then a simple effective theory which is similar to the
St\"uckelberg model. It is described in section 4.
Finally in section 5 we summarize our results  and comment on massive Yang-Mills.
Sections 2 and 3 are rather technical and awkward partly 
since even the corresponding treatment of the conventional 
case as given in \cite{Simple}--\cite{Time} is not fully developed.
One may skip
these sections in a first reading.

\setcounter{equation}{0}
\section{A simple example}
 Consider a
nonrelativistic particle in two dimensions confined to move along a straight line. A
simple Lagrangian for this system is 
\be &&L = \half(\dot{x}^2 + \dot{y}^2) -
\lambda(x\sin\theta_0 - y\cos\theta_0), \e{201}
 where $\la$ is a Lagrange multiplier
and $\theta_0$ a fixed angle. Variation with respect to $\la$ yields the constraint
$\phi_1=0$, where \be
&&\phi_1\equiv x \sin\theta_0-y\cos\theta_0,
\e{202}
which means that the particle is confined to move along a
 line through the origin with angle
$\theta_0$ relative to the $x$-axis. The corresponding Hamiltonian is
\be
&&H = \dot{x}p_x +\dot{y}p_y - L 
       = \half(p_x^2 + p_y^2) + \lambda(x\sin\theta_0 - y\cos\theta_0)
\e{203}
and in addition we have the primary constraint
\be
&&P_\la \equiv {\d L \over \d\la} = 0.
\e{204}
The total Hamiltonian is therefore $H_T=H+vP_\la$. The consistency condition
$\dot{P_\la}=\{P_\la,H_T\}=0$ etc yield then $v=0$ and the complete set of constraint
variables; $P_\la$, $\la$, $\phi_1$, $\phi_2$, where $\phi_1$ is given by \r{202} and
\be
&&\phi_2 \equiv p_x\sin\theta_0 - p_y\cos\theta_0.
\e{205}
All these constraints are of second class. 

We turn now to the quantum theory of the above system. We impose therefore the
canonical commutation relations (the nonzero part)
\be
&&[x,\,p_x]_- = [y,\,p_y]_-=[\lm,\,P_\lm]_- =i,
\e{206}
which yields
\be
&&[\phi_1,\,\phi_2]_- = i
\e{207}
for the operators $\phi_{1,2}$. The algebra of the constraint operators 
$P_\la$, $\la$, $\phi_1$, $\phi_2$ is, thus, of the form \r{1} with
$U_{ab}^{\;\;c}=0$. The form of the generalized BRST operator is,  therefore, that of
an abelian gauge theory according to the rules of \cite{RMa}. This means that we have
to insert a ghost for every constraint. We may write (a = 1,2)
\be
&&Q = \cC^a\phi_a + \bcP^1\la+ \bcP^2P_\la, 
\e{208}
where the fermionic ghost operators $\cC^a,\;\cP_b,\;\bcC_b,\;\bcP^a$ satisfy
the anticommutation relations (the nonzero part)
\be
&&[\cC^a,\,\cP_b]_+ = [\bcC_b,\,\bcP^a]_+ = \delta_b^a.
\e{209}
This BRST operator has the property
\be
&&Q^2 = i(\cC^1\cC^2 +
\bcP^1\bcP^2).
\e{2091} 
In \r{208} we have split the ghosts into two equally large
 sets, one of which we denote as
ghosts and one of which we consider as conjugate momenta to antighosts ($\bcC_a$). The
reason for this split is due to the condition \r{4}. In the ordinary case it was
found that \r{4} may always be satisfied if one considers dynamical Lagrange
multipliers and antighosts which are equal in number to the basic (usually
nontrivial) constraints and ghosts. (We have then \eg $Q=\cC\phi_1+P_\la\bcP$ which
describes a first class system.) Notice, however, that 
the split in ghosts and conjugate momenta to antighosts in \r{208} 
only is tentative. In fact, later we will find that it is not the most natural one.

In the present case we expect \r{4} to be satisfied
if the second class constraints may be split into equally large sets one of which
fixes the Lagrange multipliers. In fact, for our model \r{201} the condition \r{4} is
satisfied by expressions of the form
\be
&&\del=g^a\psi_a^{\dag},
\e{210}
where $\psi_a$, $a=1,2$, are two complex bosonic constraint operators and $g^a$,
$a=1,2$, two complex fermionic ghost operators which may be chosen to satisfy
\be
&&[g^a, g^b]=[g^a, g^{\dag b}]=0,\nn\\
&&[\psi_1, \psi_2]=[\psi_1,
\psi_1^{\dag}]=[\psi_2,
\psi_2^{\dag}]=0,\;\;\;[\psi_1,
\psi_2^{\dag}]=2i. 
\e{211}
Explicit realizations are \eg
\be
&&g^a=\half(\cC^a-i\bcP^a),\;\;\;\psi_1=\phi_1-i\la,\;\;\;\psi_2=\phi_2-iP_\la
\e{212}
or
\be
&&g^1=\half(\cC^1+i\bcP^2),\;\;\;g^2=
\half(\cC^2-i\bcP^1),\;\;\;\psi_1=\phi_1+iP_\la,\;\;\;\psi_2=\phi_2-i\la.
\e{213}
In fact, there are infinitely many ways in which \r{210} and \r{211} may be
realized. To see this one may observe that $Q$ is invariant under the scale
transformations ($\ga_1$ and $\ga_2$ are real parameters and $U_{1,2}$ unitary
operators)  \be
&U_1(\ga_1):&\phi_1\ra e^{\ga_1}\phi_1,\;\;\phi_2\ra e^{-\ga_1}\phi_2,\;\;\cC^1\ra
e^{-\ga_1}\cC^1,\;\;\cC^2\ra e^{\ga_1}\cC^2,\nn\\
&&\cP_1\ra
e^{\ga_1}\cP_1,\;\;\cP_2\ra e^{-\ga_1}\cP_2,\nn\\
&U_2(\ga_2):&\la\ra e^{\ga_2}\la,\;\;P_\la\ra e^{-\ga_2}P_\la,\;\;\bcP^1\ra
e^{-\ga_2}\bcP^1,\;\;\bcP^2\ra e^{\ga_2}\bcP^2,\nn\\
&&\bcC_1\ra
e^{\ga_2}\bcC_1,\;\;\bcC_2\ra e^{-\ga_2}\bcC_2,
\e{2131}
and under the rotations ($\theta_1$ and $\theta_2$ are real parameters and $R_{1,2}$
unitary operators)
\be
&&R_1(\theta_1):\;\;\phi_1\ra \cos{\theta_1}\phi_1+\sin\theta_1\phi_2,\;\;\phi_2\ra
\cos\theta_1\phi_2-\sin{\theta_1}\phi_1,\nn\\&&\qquad\qquad\;\;\cC^1\ra
\cos{\theta_1}\cC^1+\sin\theta_1\cC^2,\;\;\cC^2\ra
\cos\theta_1\cC^2-\sin{\theta_1}\cC^1,\nn\\
&&\qquad\qquad\;\;\cP_1\ra
\cos{\theta_1}\cP_1+\sin\theta_1\cP_2,\;\;\cP_2\ra
\cos\theta_1\cP_2-\sin{\theta_1}\cP_1,\nn\\
 &&R_2(\theta_2):\;\;\la\ra
\cos{\theta_2}\la+\sin\theta_2 P_\la,\;\;P_\la\ra\cos\theta_2 P_\la
-\sin{\theta_2}\la,\nn\\&&\qquad\qquad\;\;\bcP^1\ra
\cos{\theta_2}\bcP^1+\sin\theta_2\bcP^2,\;\;\bcP^2\ra\cos\theta_2
\bcP^2 -\sin{\theta_2}\bcP^1,\nn\\
&&\qquad\qquad\;\;\bcC_1\ra
\cos{\theta_2}\bcC_1+\sin\theta_2\bcC_2,\;\;\bcC_2\ra\cos\theta_2
\bcC_2 -\sin{\theta_2}\bcC_1.
 \e{2132}
These transformations allow us, therefore,  to represent $g^a$ and $\psi_a$ in \r{210}
in terms of four parameters starting from either \r{212} or \r{213}. (Notice that
\r{212} and \r{213} are related by an $R_2$-rotation in 90 degrees.) However, this
does not exhaust all possibilities. We may also consider representations \r{210}
in \r{4} where
\be
&&g^1=\half(\cC^1+i\cC^2), \;\;\; g^2=\half(\bcP^1+i\bcP^2),\nn\\
&&\psi_1=\phi_1+i\phi_2,\;\;\; \psi_2=\la+iP_\la.
\e{214}
Instead of \r{211} $\psi_a$ satisfy here $[\psi_a, \psi_b^{\dag}]=2\del_{ab}$. Again
the transformations \r{2131} and \r{2132} allow us to represent 
$g^a$ and $\psi_a$ in \r{210} in terms of four parameters but now starting from
\r{214} which is different from the previous representation.

Using \r{4} with the representation \r{210} the  solutions with nonzero norms of the
conditions \r{5} may be shown to satisfy 
 \be
&&i)\ g^a|{ ph}\rangle = 0,\quad \psi_a|{ ph}\rangle = 0,
\e{215}
or
\be
&& ii)\ g^{a\dagger}|{ ph}\rangle = 0, \quad \psi_a^\dagger| ph\rangle = 0.
\e{216}
Both these sets of solutions should be physically equivalent. (Notice that 
 the first conditions in \r{215} and
\r{216} by themselves imply
$Q^2|{\rm ph}\rangle = 0$.)

We shall now show that the nontrivial solutions of \r{5}, \ie the solutions up to
zero norm states, may be represented in the form \be
&&|s\hb=e^\Lambda|\phi\hb,
\e{217}
where $\Lambda$ is an Hermitian operator and $|\phi\hb$ a simple state determined by
Hermitian conditions. In the ordinary case we have the representation \r{217} where
$|s\hb$ is a BRST singlet, $\Lambda=[Q, \psi]$ where $\psi$ is a gauge fixing
fermion, and where  $|\phi\hb$ is determined by a complete irreducible set of BRST
doublets in involution \cite{BRM}. Since the BRST charge is no longer nilpotent in this
generalized BRST case, operators and states are decomposed into higher multiplets than
doublets. For our simple model the basic variables may be decomposed into BRST
triplets. In fact, we have the following elementary hermitian BRST triplets
\be
\begin{array}{llcl}
   (1)&\{\cP_1,\,\phi_1,\,\cC^2\}, & \quad (2)&\{\cP_2,\,\phi_2,\,\cC^1\},\\
   (3)&\{\bcC_2,\,P_\lm,\,\bcP^1\}, & \quad  (4)&\{\bcC_1,\,\lm,\,\bcP^2\},
   \end{array}
\e{218}
where the entries $(a,b,c)$ are related by $b\propto[Q,a]$, $c\propto[Q,b]$.
All these triplets form generalized quartets or rather sixtets. However, half of them
are in involution. In fact the following four combinations of two of the above
triplets are in involution: \be
    [1] : (1) + (3), \qquad [2] : (2) + (4), \qquad
    [3] : (1) + (4), \qquad [4] : (2) + (3).
\e{219}
The most simple   BRST invariant solutions of the form \r{217} should therefore be
\be
&&|s\hb_l=e^{\Lambda_l}|\phi\hb_l, \quad l=1,2,3,4,
\e{220}
where $|\phi\hb_l$ is a state determined by the operators in the combination $[l]$ in
\r{219} \ie \be
&& \cP_1|\phi\rangle_1 = \phi_1|\phi\rangle_1 = \cC^2|\phi\rangle_1
     = \bcC_2|\phi\rangle_1 = P_\la|\phi\rangle_1 = \bcP^1|\phi\rangle_1 = 0,\nn\\
&&    \cP_2|\phi\rangle_2 = \phi_2|\phi\rangle_2 = \cC^1|\phi\rangle_2
     = \bcC_1|\phi\rangle_2 = \la|\phi\rangle_2 = \bcP^2|\phi\rangle_2 = 0,\nn\\
&&\cP_1|{\phi}\rangle_3 = \phi_1|{\phi}\rangle_3
      = \cC^2|{\phi}\rangle_3 = \bcC_1|{\phi}\rangle_3
      = \la|{\phi}\rangle_3 = \bcP^2|{\phi}\rangle_3 = 0,\nn\\
    &&\cP_2|{\phi}\rangle_4 = \phi_2|{\phi}\rangle_4
      = \cC^1|{\phi}\rangle_4 = \bcC_2|{\phi}\rangle_4
      = P_\la|{\phi}\rangle_4 = \bcP^1|{\phi}\rangle_4 = 0.
\e{221}
If we choose
\be
&&\Lambda_1= \phi_2\lm -i\bcP^2\cP_2 + i\cC^1\bcC_1,\nn\\
&&\Lambda_2= - \phi_1P_\lm -i\bcP^1\cP_1 + i\cC^2\bcC_2,\nn\\
&&\Lambda_3= \phi_2P_\lm +i\bcP^1\cP_2 + i\cC^1\bcC_2,\nn\\
&&\Lambda_4= \phi_1\lm - i\cC^2\bcC_1 - i\bcP^2\cP_1,
\e{222}
in \r{220} then $|s\hb_{1,2}$ satisfy \r{215} in the realization \r{212} and 
$|s\hb_{3,4}$ satisfy \r{215} in the realization \r{213}. Notice that
$[Q,\Lambda_l]=0$, $Q|\phi\hb_l=0$, which implies $Q|s\hb_l=0$. Notice also that the
conditions \r{221} are invariant under the scale transformations \r{2131} but not
under the rotations \r{2132} except for rotations in 180 degrees. Furthermore, they are
related by the rotations \r{2132}. Starting from 
\eg $|s\hb_1=e^{\Lambda_1}|\phi\hb_1$ one
may by successive rotations in 90 degrees reach all the four cases above. Arbitrary
rotations yield singlets which interpolate  the ones in \r{220}. The unitary scale
transformations \r{2131} scale $\Lambda_l$ according to  
\be &&\Lambda_l\;\ra\;\al_l
\Lambda_l,\;\;\;\al_1=e^{-\ga_1+\ga_2},\;\;\;\al_2=e^{\ga_1-\ga_2},
\;\;\;\al_3=e^{-\ga_1-\ga_2},\;\;\;\al_4=e^{\ga_1+\ga_2}.
 \e{223}
 Now, in
distinction to the ordinary BRST case these unitary scale transformations, $U_1$ and
$U_2$, do not yield unity on the $|\phi\hb_l$-states. Instead we have \be
&&|\phi\hb_l\;\ra\;{1\over \sqrt{\al_l}}|\phi\hb_l.
 \e{224}
Unitary equivalent solutions to \r{220} may therefore be written in the following
 form \be
&&|s\hb_l={1\over \sqrt{\al_l}}e^{\pm\al_l \Lambda_l}|\phi\hb_l, \quad l=1,2,3,4,
\e{225}
where $|\phi\hb_l$ is the same solution of \r{221} as in \r{220}. (Notice,
however, that $|\phi\hb_l$ scaled with any constant still satisfy \r{221}.) Here as
well as in the ordinary BRST case $|\phi\hb_l$ is not an inner product space. However,
while $|\phi\hb_l$ in the ordinary case may be regularized to an inner product state
it is here regularized to a zero norm state which is the reason why the property
\r{224} is allowed for unitary transformations here. The reason behind this difference
compared to the ordinary case is that our scheme in a sense makes use of an
unnecessary large number of ghost variables for second class constraints. (A scheme
with a nilpotent charge like the one in \cite{BTL1,BTL2} should not have the property
\r{224}.) Obviously, the measure operator $e^{\Lambda_l}$ 
acts here in a more nontrivial way
as compared to the ordinary case since a scale 
transformation of $\Lambda_l$ affects the
norms. (In the ordinary case $e^{\Lambda_l}$ is just a regulator.)

The above properties may be elaborated further. 
The $\Lambda_l$ operators in \r{222} satisfy
the properties
\be
&&\Lambda_l|\phi\hb_{l'}=0\;\;\;{\rm for}\;\;\;l\neq l' .
\e{2251}
Furthermore, $\Lambda_1$ and $\Lambda_2$ as well as 
$\Lambda_3$ and $\Lambda_4$ satisfy a closed algebra.
Consider the following two identifications
\be
&(1)& K_1 = \Lambda_1,\quad K_2 = \Lambda_2,\quad K_3 =
 [\Lambda_1, \Lambda_2],
\e{2252}
\be
&(2)& K_1 = \Lambda_3,\quad K_2 = \Lambda_4,\quad K_3 = 
[\Lambda_3, \Lambda_4].
\e{2253}
For both cases, (1) and (2), we have then
\be
&&K_3 =[K_1, K_2],\quad [K_3, K_1] = 2K_1,\quad [K_3, K_2] = -2K_2,
\e{2254}
which is an SL(2,R) algebra. The $K_3$ operator in the two cases satisfies the
properties
\be
&(1)& K_3|\phi\hb_1=-|\phi\hb_1,\;\;\;
K_3|\phi\hb_2=|\phi\hb_2,\;\;\;K_3|\phi\hb_{3,4}=0,
 \e{2255}
\be
&(2)&  K_3|\phi\hb_{1,2}=0,\;\;\;K_3|\phi\hb_3=-|\phi\hb_3,\;\;\;
K_3|\phi\hb_4=|\phi\hb_4.
\e{2256}
By means of these properties we may construct the following more general expressions
of the solutions \r{220} (cf. the corresponding constructions for general gauge
theories in \cite{GRM}): \be
&&|s\hb_{1,2}=e^{\al \Lambda_1+\beta \Lambda_2}|\phi\hb_{1,2},
\qquad
|s\hb_{3,4}=e^{\al \Lambda_3+\beta \Lambda_4}|\phi\hb_{3,4}.
\e{2257}
These expressions may be reduced to the form \r{225} by means of the following
factorizations
\be
&& e^{\al K_1+\beta K_2}
          = e^{\del_1 K_1}e^{\del_2 K_2}e^{\del_3 K_3}
          = e^{\del'_2 K_2}e^{\del'_1 K_1}e^{\del'_3 K_3},
\e{2258}
where 
\be
&& e^{-\del_3} = \cosh\sqrt{\al\beta},\quad
    \del_1 = \al{\tanh\sqrt{\al\beta} \over \sqrt{\al\beta}},\quad
    \del_2 = \half\beta{\sinh2\sqrt{\al\beta} \over \sqrt{\al\beta}}
\e{22590}
and
\be
&&e^{\del'_3} = \cosh\sqrt{\al\beta},\quad
    \del'_1 = \half\al{\sinh2\sqrt{\al\beta} \over \sqrt{\al\beta}},\quad
    \del'_2 = \beta{\tanh\sqrt{\al\beta} \over \sqrt{\al\beta}}
\e{22591}
for $\al\beta>0$. For $\al\beta<0$ we have the same expressions with the
corresponding trigonometric functions and $\sqrt{\al\beta}$ replaced by 
$\sqrt{-\al\beta}$. Eq.\r{2258} inserted into \r{2257}  
yields by means of \r{2251}, \r{2255} and
\r{2256}:
\be
&&|s\hb_1=\la e^{\del\al \Lambda_1}|\phi\hb_1,\;\;\;
|s\hb_2=\la e^{\del\beta
\Lambda_2}|\phi\hb_2,\nn\\
&&|s\hb_3=\la e^{\del\al \Lambda_3}|\phi\hb_3,\;\;\;
|s\hb_4=\la e^{\del\beta
\Lambda_4}|\phi\hb_4,
\e{22592}
where
\be
&&\la=\cosh\sqrt{\al\beta},\;\;\;\del={\tanh\sqrt{\al\beta}\over\sqrt{\al\beta}}.
\e{22593}
Thus, in order for the expressions \r{2257} to have the same normalizations as \r{225}
when using the same $|\phi\hb_l$-states we have to multiply $|s\hb_{1,3}$ and
$|s\hb_{2,4}$ in
 \r{2257} by the factor $\sqrt{{2\sqrt{\beta/\al}/\sinh2\sqrt{\al\beta}}}$ and
$\sqrt{{2\sqrt{\al/\beta}/\sinh2\sqrt{\al\beta}}}$ respectively.  These results may
be directly compared with the corresponding results for general gauge theories
\cite{GRM}. The main difference is that for general gauge theories we have
$K_3|\phi\hb_l=0$ and $\la=1$. (Notice that $K_i$ in \cite{GRM} are different from
ours. To compare one has here to make the replacement $K_1\ra-K_2,\;K_2\ra K_1$
implying $\al\ra-\beta,\;\beta\ra\al$.)

In order to determine the time evolution of the BRST invariant solutions we have to
determine an appropriate BRST invariant Hamiltonian. 
The original Hamiltonian operator is
\be
  &&  H = \half(p_x^2 + p_y^2) + \lm\phi_1,
\e{226}
which is not BRST invariant since
\be
&&   [Q,\,H] = i(\cC^1\phi_2 - \cC^2\lm - \phi_1\bcP^2).
\e{227}
A BRST invariant Hamiltonian may, however, be obtained by adding ghosts
to $H$. We have then
\be
&&   H_{\rm BRST} = H + i\cP_1\bcP^2 + i\bcC_1\cC^2 - i\cP_2\cC^1, 
\e{228}
which is one possible candidate of a BRST invariant Hamiltonian.
Other candidates are obtained by truncation. We have
\be
   H'_{\rm BRST} = \half(p_x^2 + p_y^2) - i\cP_2\cC^1
\e{229}
by first removing the $\la\phi_1$-term in the original Hamiltonian \r{226}.
Furthermore, since
\be
&& \half(p_x^2 + p_y^2) = \half(p_x\cos\theta_0 + p_y\sin\theta_0)^2
                        + \half(\phi_2)^2
\e{230}
we have also the following completely ghost independent BRST invariant Hamiltonian 
\be
&&
   H''_{\rm BRST} = \half(p_x\cos\theta_0 + p_y\sin\theta_0)^2.
\e{231}

A time dependent BRST invariant solution is naturally expected to be given by
\be
&&|s,t\hb=e^{-iH_0t}|s\hb,
\e{232}
where $H_0$ is one of the the BRST  invariant Hamiltonians 
\r{228}, \r{229} or \r{231}.
Which one should be chosen? 
From the results obtained for ordinary BRST quantization
\cite{Path}  the appropriate Hamiltonian should be such that we also have
$|\phi, t\hb=e^{-iH_0t}|\phi\hb$. This requires $H_0$ 
to commute with $\Lambda_l$. We notice
then that $H_{\rm BRST}$ in \r{228} does not commute with any 
$\Lambda_l$.  $H'_{\rm BRST}$
in \r{229} commutes with $\Lambda_1$ and $\Lambda_3$ but not with
 $\Lambda_2$ and $\Lambda_4$, while
only $H''_{\rm BRST}$ in \r{231} commutes with all $\Lambda_l$. Notice also that only 
$H''_{\rm BRST}$ is invariant under the scaling \r{2131} and under the rotations
\r{2132}.
 $H''_{\rm BRST}$ seems therefore to be the appropriate choice for $H_0$. 
(All this should be viewed from the corresponding 
reparametrization invariant theory whose BRST treatment allows for transformations 
between the Hamiltonian and the constraints \cite{Time}.)

In order to be able to relate the above results to path integral formulations we
have to determine appropriate quantization rules. From \r{211},\r{215} and \r{216} we
conclude that the unphysical part of the original state space is spanned by oscillator
variables satisfying the algebra (the nonzero elements) 
\be &&|g^a,
k_b^{\dag}]_+=\del^a_b,\qquad[A, B^{\dag}]_-=1, 
\e{233}
where $g^a$, $k_a$, $a=1,2$, are fermionic and $A$ and $B$ bosonic. $g^a$ may be
identified with $g^a$ in \r{210}, and $A$ and $B$ with $\psi_1$($\psi_2$) and $\half
i \psi_2$($-\half i \psi_1$) in \r{211}. The corresponding Fock representation consists
when diagonalized of half positive and half indefinite metric oscillators both for
bosons and fermions separately. This implies in turn that half the hermitian unphysical
operators have imaginary eigenvalues \cite{Proper}. For the bosons, $x$ and $y$,
we should have real eigenvalues since they are treated on equal level which implies
that the Lagrange multiplier $\la$ should have imaginary eigenvalues. Since the BRST
invariant Hamiltonian $H_0$ should have  real eigenvalues 
$H_0=H_{BRST}$ is excluded. For the fermions a natural choice is such that  $\Lambda_l$ has
imaginary eigenvalues.  This implies
that the ghost variables $\cC^a$ and $\bcC_a$ should have opposite properties, one
should have  real eigenvalues and the other imaginary ones.

Relation   \r{232} and the form  \r{2257} for the solutions lead to the following 
transition amplitudes ($H_0$ and $\Lambda_l$ commute)
\be &&\vb s', t'|s, t \hb=\vb s'|e^{iH_0(t'-t)}|s\hb=
\left\{\begin{array}{c}_l\vb\phi'|e^{iH_0(t'-t)+\al \Lambda_1+\beta
\Lambda_2}|\phi\hb_l,\;\;\;l=1,2\\ _l\vb\phi'|e^{iH_0(t'-t)+\al 
\Lambda_3+\beta
\Lambda_4}|\phi\hb_l,\;\;\;l=3,4.\end{array}\right.
 \e{234}
Replacing $\al$ and $\beta$ by $\al\ra(t'-t)\al$ and $\beta\ra(t'-t)\beta$ we are
then  led to the path integral representation (cf \cite{Path}) \be
&&\vb s', t'|s,t\hb=\int dq' dq {\phi'}^*(q^{'*})\phi(q)\vb q', t'|q^*,
t\hb,\nn\\
&&\vb q', t'|q^*,
t\hb=\int[dq][dp] e^{i\int_t^{t'}(p\dot{q}-H_{\rm eff})dt},
\e{235}
where $q$ denotes collectively all the coordinates ($x, y, \la, \cC^a, \bcC_a$) and
$p$ their corresponding conjugate momenta. According to the previous results
the normalization of these time dependent solutions is also time dependent. In order
to have time independent normalization \r{234} has to be multiplied by the
factor ${1/\sinh2(t'-t)\sqrt{\al\beta}}$. Notice that \r{221} are conditions which
act like boundary conditions in the path integral. $H_{\rm eff}$ is an effective
pseudoclassical Hamiltonian. For $H_0=H''_{\rm BRST}$ we have the following choices
\be &&l=1,2:\quad H_{\rm eff}= 
\half(p_x\cos\theta_0 + p_y\sin\theta_0)^2+\al \la_1+\beta
\la_2,\nn\\ &&l=3,4:\quad H_{\rm eff}= \half(p_x\cos\theta_0 + p_y\sin\theta_0)^2 +\al
\la_3+\beta \la_4,
\e{236} 
where  $\la_l$ are real pseudoclassical expressions from $\Lambda_l$. 
The choice of effective Hamiltonian depends on the choice of boundary condition.
In fact, by means of the rotations \r{2132} we may also obtain $H_{\rm eff}$ and
boundary conditions 
 which interpolate between the above choices. One may   convince oneself
that for any  choice of an allowed effective Hamiltonian in \r{235} together with
the corresponding boundary condition implied by \r{221} and its interpolations by
means of \r{2132} we get one and the same result: Integrations over the Lagrange
multiplier and its conjugate momentum produce
 delta functions 
so that we for any choice effectively have
\be &&\vb x', y', t'|x, y, t\hb=\nn\\
&&=\int[dx][dy][dp_x][dp_y]
\del(\phi_1)\del(\phi_2)e^{i\int_t^{t'}\left(p_x\dot{x}+p_y\dot{y}-\half(p_x\cos\theta_0
+ p_y\sin\theta_0)^2\right)}, 
\e{238}
 which is in agreement with the prescription of
\cite{PSe}.

Now the above effective Hamiltonians are not the natural ones from a geometrical
point of view. They are also not ``gauge fixed" in the sense   that the
corresponding Lagrangians are not regular. In order to find out what a more natural
effective Hamiltonian requires we notice now the following relations between the
Hamiltonians \r{228}, \r{229} and \r{231}.  
\be
&&H_{BRST}=H'_{BRST}+\Lambda_4=H''_{BRST}+\half i
[\Lambda_3,\Lambda_1]+\Lambda_4.
\e{239}
Thus, it seems that in order to have an effective
Hamiltonian corresponding to \r{228} we must consider solutions of the following
form
\be
&&|s\hb=e^{\al\left(\Lambda_4+\half i
[\Lambda_3, \Lambda_1]\right)}|\phi\hb_4.
\e{240}
However, since $\half i[\Lambda_3, \Lambda_1]$ has real
 eigenvalues and $\Lambda_4$ imaginary ones
this leads to a complex effective Hamiltonian in the path integral. Furthermore, the
corresponding effective Lagrangian is not regular.

Now the forms \r{220} and \r{2257}  of the solutions are not the only ones. By means
of the representation \r{214} in \r{210} we have also 
\be
&&|s\hb_l=e^{B_l}|\phi\hb_l,
\e{241}
where
\be
&&B_1=\al\half i[\Lambda_1, \Lambda_3]-\beta\half i
[\Lambda_1,\Lambda_4],\;\;\;B_2=\al\half i[\Lambda_2,
\Lambda_3]+\beta\half i[\Lambda_2,\Lambda_4],\nn\\ &&B_3=-
\al\half i[\Lambda_3, \Lambda_1]+\beta\half
i[\Lambda_3,\Lambda_2],\;\;\;B_4=\al\half i[\Lambda_4, \Lambda_1]-
\beta\half i[\Lambda_4,\Lambda_2], 
\e{242}
where in turn $\al$ and $\beta$ are real {\em positive} parameters.
The scaling transformations \r{2131} changes $\al$ and $\beta$ to arbitrary positive
values while a rotation \r{2132} in 180 degrees does not change the sign of $B_l$. The
particular signs of $\al$ and $\beta$ in \r{242} is directly related to our choice to
let $x$ and $y$ represent positive metric states and $\la$ negative metric ones. By
means of
the particular solution
\be
&&|s\hb=e^{B_3}|\phi\hb_3, \;\;\;B_3=-\al\half 
i[\Lambda_3, \Lambda_1]+\al\half
i[\Lambda_3,\Lambda_2],
\e{243}
we may now define transition amplitudes for imaginary times by ($\tau'>\tau$)
\be
&&\vb s',\tau'|s,\tau\hb\equiv\vb s'|e^{-H_0(\tau'-\tau)}|s\hb=\,_3\vb\phi'|
e^{-H_0(\tau'-\tau)-2\al\half i[\Lambda_3, \Lambda_1]+2\al\half
i[\Lambda_3,\Lambda_2]}|\phi\hb_3,
\e{244}
where $H_0=H''_{BRST}$ which commutes with $B_l$.
Replacing $2\al$ by $\tau'-\tau$ we get the path integral expression \r{235} with
imaginary times with the effective BRST invariant Hamiltonian
\be
&&H_{\rm eff}=\half(p_x^2+p_y^2+P_\la^2)-i\cP_2\cC^1+i\bcP^1\bcC_2,
\e{245}
which is a natural geometric and gauge fixed Hamiltonian. 
In fact, the corresponding effective Lagrangian may be written
\be
&&L_{\rm eff}=\half(\dot{x}^2+\dot{y}^2+\dot{\la}^2)+i\dot{\bcC}_a\dot{\cC}^a,
\e{246}
where we have made the following ghost redefinition: 
$\cC^1\,\ra\,\bcP^2$, $\bcC_2\,\ra\,\cP_1$.
Notice 
that the conditions
on $|\phi\hb_3$ in \r{221} lead to the boundary conditions  $\phi_1=0$ and $\la=0$
etc. in \r{235}. 

 \setcounter{equation}{0} \section{Application to Massive Vector Fields}
Encouraged by the previous results we treat now  the massive vector field whose
Lagrangian   is given by  (we use a space-like Minkowski metric)
\be
&&\cL=- {1 \over 4}F_{\mu\nu}F^{\mu\nu} - {1 \over 2}m^2 A_{\mu} A^{\mu},
\e{300}
where $F_{\mu\nu} = \partial_\mu A_{\nu} -
\partial_\nu A_{\mu }$. This Lagrangian leads to two second class constraints which
eliminates $A^0$. However, these constraints do not have the necessary structure
for a BRST quantization on an inner product space
as described in the previous section. 
For that to be the case we need an
additional unphysical degree of freedom. Since Lorentz covariance will require $A^0$
to be quantized with negative metric states we must add a
  bosonic unphysical
scalar field, $\varphi$, which then must be  quantized with
 positive metric states in order to be
able to obtain inner product solutions. Since, we do not know off hand how this
scalar should couple in the Lagrangian we choose for simplicity the most
trivial coupling possible which leads to a constraint algebra like
the one we had for the previous simple model. 
We consider therefore the following Lagrangian
\be
   &&\cL = - {1 \over 4}F_{\mu\nu}F^{\mu\nu} - {1 \over 2}m^2 A_{\mu} A^{\mu}-{1
\over 2}m^2 \varphi^2,
        \e{301}
where $\varphi$ is a real scalar field. The conjugate
momenta to $A_{\mu }$ and $\varphi$ are given by
\be
   &&\pi^{\mu} = {\partial\cL\over \partial\partial_0A_{\mu }}=F^{\mu 0},\;\;\;\pi=
{\partial\cL\over\partial\partial_0\varphi}=0.
  \e{302}
Hence, we have the primary constraints
\be
&&\phi_1\equiv\pi^{0} =0,\;\;\;P_\la\equiv\pi=0.
\e{303}
The total Hamiltonian  is 
  \be
  &&H_T = \int\cH_T(x) d^{3}x,\quad \cH_T(x) =\cH_0(x) + v(x)\pi^0(x)+w(x)\pi(x), 
  \e{304}
where $\cH_0(x)$ is the canonical energy density (up to a divergence)
  \be
   &&\cH_0 =  \half \pi_{i}\pi^{i} + A^{0}\partial_{i}\pi^{i}
         + {1 \over 4}F_{ij}F^{ij} + \half m^2 A_{\mu} A^{\mu}+\half m^2 \varphi^2.
  \e{305}
The total Hamiltonian may be used to find out the consequences of  the consistency
conditions $\partial_0{\pi}^{0} = 0$ and $\partial_0{\pi} = 0$ for the primary
constraints. 
By means of the Poisson bracket defined by
\be
&&\left.\{A^\mu(x),
\pi_{\nu}(y)\}\right|_{x^0=y^0}=\del^\mu_\nu\del^3(x-y),\nn\\
&&\left.\{\varphi(x),
\pi(y)\}\right|_{x^0=y^0}=\del^3(x-y),
\e{306}
we find
\be
   &&\partial_0{\pi}^{0} = \{\pi^{0},\,H_T\} = -\partial_{i}\pi^{i} +
m^2A^{0},\nn\\
&&\partial_0{\pi} = \{\pi,\,H_T\} = -
m^2\varphi.
   \e{307}
Thus, consistency requires us to impose the secondary constraints
\be \phi_2\equiv-\partial_{i}\pi^{i} +
m^2A^{0}=0,\;\;\;\la\equiv m^2\varphi=0.
\e{308}
The further consistency conditions $\partial_0{\phi}_2 = 0$ and $\partial_0{\la} = 0$
require in turn that $v=\partial_iA^i$ and $w=0$ in the total Hamiltonian \r{304}.
The last term in \r{301} may obviously be removed without affecting the theory at the
classical level. However, only with  a term like this present do we have a constraint
structure in exact correspondence to what we had 
for the simple model in the previous
section and which is necessary for an inner product quantization. (The factor $m^2$
may be chosen arbitrary as will be shown in the next section.) The equal-time Poisson
algebra of the constraints 
\r{303} and
\r{308} are here \be
   && \{\phi_{1}(x),\,\phi_{1}(y)\} =
\{\phi_{2}(x),\,\phi_{2}(y)\}=\{P_\la(x),\,P_\la(y)\} = \{\la(x),\,\la(y)\}=0,\nn\\
&&\{\phi_{1}(x),\,\phi_{2}(y)\} =\{\la(x),\,P_\la(y)\} = 
m^2\delta^{3}(x-y). 
\e{309}

We turn now to the quantum theory. From \r{306} we have the basic nonzero
equal-time commutators
\be
&&[A^\mu(x),
\pi_{\nu}(y)]=i\del^\mu_\nu\del^3(x-y)\nn\\
&&[\varphi(x),
\pi(y)]=i\del^3(x-y),
\e{310}
which lead to 
\be
&&[\phi_{1}(x),\,\phi_{2}(y)] =[\la(x),\,P_\la(y)] = i
m^2\delta^{3}(x-y) 
\e{311}
for the constraint operators. In analogy with \r{208} we define the BRST operator to
be
\be
&&Q=\int (\cC^a\phi_a+\bcP^1\la+\bcP^2P_\la)d^3x\equiv\nn\\
&&\equiv\int\left\{\cC^1\pi^0-\cC^2(\partial_{i}\pi^{i} -
m^2A^{0})+\bcP^1m^2\varphi+\bcP^2\pi\right\}d^3x,
\e{312}
where we have introduced the fermionic ghost field operators $\cC^a$, $\cP_b$ and 
$\bcC_a$, $\bcP^b$ satisfying
\be
&&[\cC^a(x), \cP_b(y)]=\del^a_b\del^3(x-y),\;\;\;[\bcC_a(x),
\bcP^b(y)]=\del^b_a\del^3(x-y).
\e{313}
We have then
\be
&&Q^2=im^2\int(\cC^1\cC^2+\bcP^1\bcP^2)d^3x.
\e{314}
In analogy with the previous section we may \eg define 
BRST invariant states    by
\be
&&|s\hb_{1,2}=\exp{\left(\int(\al(x)\Lambda_1(x)+\beta(x)
\Lambda_2(x))d^3x\right)}|\phi\hb_{1,2}\nn\\
&&|s\hb_{3,4}=\exp{\left(\int(\al(x)\Lambda_3(x)+\beta(x)
\Lambda_4(x))d^3x\right)}|\phi\hb_{3,4},
\e{315}
where $|\phi\hb_{l}$ satisfy the local field conditions corresponding to \r{221}.
$\al(x)$ and $\beta(x)$  are parameter functions, and $\Lambda_l(x)$ are the
following BRST invariant operator expressions 
\be
&&\Lambda_1(x)= \phi_2\lm -im^2\bcP^2\cP_2 + im^2\cC^1\bcC_1,\nn\\
&&\Lambda_2(x)= - \phi_1P_\lm -im^2\bcP^1\cP_1 + im^2\cC^2\bcC_2,\nn\\
&&\Lambda_3(x)= \phi_2P_\lm +im^2\bcP^1\cP_2 + im^2\cC^1\bcC_2,\nn\\
&&\Lambda_4(x)= \phi_1\lm - im^2\cC^2\bcC_1 - im^2\bcP^2\cP_1.
\e{316}
With appropriate choices of $|\phi\hb_l$ the norms of $|s\hb_l$ 
in \r{315} are finite. However, again the choices \r{315} are 
neither natural 
from a geometrical point of view nor does it lead to a regular 
effective theory.

A natural choice of  a BRST invariant Hamiltonian is 
 the following truncated version of the
 total Hamiltonian \r{304},
\be 
&&\cH'(x)=  \half \pi_{i}\pi^{i} 
         + {1 \over 4}F_{ij}F^{ij} + \half m^2 A_{i} A^{i}-\partial_{i}\pi^{0}A^{i}+
{1\over 2m^2}\partial^i\pi^0\partial_i\pi^0.
\e{317}
It commutes with all constraint operators and is therefore 
BRST invariant. If we then
choose a BRST invariant solution of the form
 \be
&&|s\hb=e^{-\al B_1}|\phi\hb_1,
\e{318}
where $\al$ is a positive constant and where
\be
&&B_1\equiv{i\over 2m^4}\left[\int  \Lambda_1(x) d^3x, \int
(\Lambda_3(y)+\Lambda_4(y)) d^3y\right]=\nn\\ &&=\int \left\{-{1\over
2m^2}(\partial_{i}\pi^{i} - m^2A^{0})^2+      \half
m^2\varphi^2-i\cC^1\cP_2-i\bcP^2\bcC_1\right\} d^3x,
\e{319}
we may  define the transition amplitude 
for imaginary times by ($\tau'>\tau$)
\be
&&\vb s',\tau'|s,\tau\hb\equiv\vb s'|e^{-H'(\tau'-\tau)}|s\hb=\,_1\vb\phi'|
e^{-H'(\tau'-\tau)-2\al B_1}|\phi\hb_1.
\e{320}
Choosing $2\al=\tau'-\tau$ we may then extract an
 effective Hamiltonian in the path integral which  contains
 the terms of the
total Hamiltonian \r{304}. However, it will also contain two additional terms. One term
is $\partial_{i}\pi^{0}\partial^{i}\pi^{0}/2m^2$ 
which is a  ``gauge fixing"
term that makes the Hamiltonian regular. The other term is 
$-(\partial_{i}\pi^{i})^2/2m^2$ which is a genuine physical term which commutes
with the BRST charge. The latter should not be in the effective Hamiltonian.
Fortunately, this term is easily removed by replacing $\cH'$ in \r{317} by
\be
&&\cH''=\cH'+(\partial_{i}\pi^{i})^2/2m^2,
\e{321}
which also is BRST invariant and commutes with all constraints. 

In a sense the construction given above is as successful as the corresponding
construction for the simple model in the previous section. However, we notice that
\eg the term $\partial_{i}\pi^{0}\partial^{i}\pi^{0}/2m^2$ is not  
an appropriate term for
an effective Hamiltonian since it does not lead to 
a ``gauge fixing" term in the corresponding effective Lagrangian which is
manifestly Lorentz invariant. In order to obtain such a Lagrangian we have to 
remove the term $\partial_{i}\pi^{0}\partial^{i}\pi^{0}/2m^2$
 and insert new terms required by manifest Lorentz invariance.
These terms should then only involve the constraint variables. 
The simplest choice is then to add the terms   
$-(\pi^0)^2/2+\pi^2/2+\partial_i\varphi\partial^i\varphi/2$.
 This may be performed in a BRST
 invariant way by just replacing $B_1$ in \r{319} by $B'_1$ given by
\be
&&B'_1=B_1+\int \left\{-{1\over 2m^2}\partial_{i}\pi^{0}\partial^{i}\pi^{0}
-\half(\pi^0)^2+\half\pi^2+\half\partial_i\varphi\partial^i\varphi
+\right.\nn\\&&\left.+im^2\cC^2\cP_1+im^2\bcP^1\bcC_2+i\partial_i\cC^2\partial^i\cP_1
-{i\over m^2}\partial_i\bcP^2\partial^i\bcC_1\right\} d^3x.
\e{322}
It is remarkable that the additional ghost terms which are required 
by BRST invariance also lead to a manifestly Lorentz 
invariant expression for the ghost terms in the effective 
Lagrangian. The final effective
 theory is described below.

 \setcounter{equation}{0} 
\section{Effective Theory for Massive Vector Fields}
Our manifestly Lorentz invariant effective
 Lagrangian for the massive vector field is given by
\be
  &\cL =& - {1 \over 4}F_{\mu\nu}F^{\mu\nu} - 
{1 \over 2}m^2 A_{\mu} A^{\mu}-\half(\d_\mu A^\mu)^2-
\half\d_\mu \varphi\d^\mu \varphi-{1
\over 2}\mu^2 \varphi^2+\nn\\&&+i\d_\mu\bcC_1\d^\mu\cC^1+im^2\bcC_1\cC^1+
i\d_\mu\bcC_2\d^\mu\cC^2+i\mu^2\bcC_2\cC^2,
        \e{401}
where we have redefined the ghost 
fields as compared to the previous section (see below). 
This Lagrangian yields the simple equations
\be
&&(\Box-m^2)A^\mu=(\Box-m^2)\cC^1=(\Box-m^2)\bcC_1=0,\nn\\
&&(\Box-\mu^2)\varphi=(\Box-\mu^2)\cC^2=(\Box-\mu^2)\bcC_2=0.
\e{402}
Thus, the five bosonic and four fermionic degrees of freedom all 
satisfy a  massive free equation. The Lagrangian \r{401} is
 quasi-invariant under the generalized BRST transformation
\be
&&s A^\mu=-\d^\mu\cC^1,\quad s\bcC_1=-i\d_\mu A^\mu,\nn\\
&&s\varphi=-m\cC^2,\quad s\bcC_2=-im\varphi,\quad s\cC^a=0.
\e{403}
which  only is nilpotent on the matter 
fields $A^\mu$ and $\varphi$. In fact, we have
\be
&&s^2A^\mu=s^2\varphi=0,\quad s^2\bcC_1=
i\Box \cC^1\approx im^2\cC^1,\quad s^2\bcC_2=im^2\cC^2.
\e{404}
The conserved current density corresponding 
to the above BRST invariance  is
\be
&&J^\mu=F^{\mu\nu}\d_\nu\cC^1-
m^2A^\mu\cC^1+(\d_\nu A^\nu)\d^\mu\cC^1+
m\varphi\stackrel{\lra}{\d^\mu}\cC^2.
\e{405}

In order to demonstrate the equivalence with what we had in
 the previous section we perform a transition to the corresponding
 Hamiltonian formulation. We define then the conjugate
momenta to the field variable in \r{401} by
\be
&&\pi^{\mu} = {\partial\cL\over \partial\partial_0A_{\mu }}=
F^{\mu 0}+\d_\nu A^\nu\eta^{\mu 0},\quad\pi=
{\partial\cL\over\partial\partial_0\varphi}=\partial^0\varphi,\nn\\
&&\cP_a=i{\partial\cL\over\partial\partial_0\cC^a}=-\partial_0{\bcC}_a,\quad
\bcP^a=i{\partial\cL\over\partial\partial_0\bcC_a}=\partial_0{\cC}^a,
  \e{406}
where $\eta^{\mu\nu}$ is the space-like Minkowski metric.
The Hamiltonian density is given by
\be
&\cH=&\pi^\mu\d_0A_\mu+\pi\d_0\varphi+i\cP_a\d_0\cC^a+i\bcP^a\d_0\bcC_a-\cL
=\half\pi^i\pi_i-\half(\pi^0)^2+\nn\\
&&+
{1\over 4}F^{ij}F_{ij}+\half m^2A^\mu A_\mu-\pi^i\d_i A^0+\pi^0\d_i
A^i+\half\pi^2+\half\mu^2\varphi^2+\nn\\ &&+\half\d_i\varphi\d^i\varphi+
i\cP_a\bcP^a-i\d_i\bcC_a\d^i\cC^a-im^2\bcC_1\cC^1-i\mu^2\bcC_2\cC^2.
\e{407} 
From \r{405} we find the  BRST charge 
\be
&&Q=\int J^0(x)d^3x=\int\left\{(\d_i\pi^i-m^2A^0)\cC^1-\pi^0\bcP^1+
m\varphi\bcP^2-m\pi\cC^2\right\}d^3x,
\e{408}
which  generates the BRST transformation \r{403} in terms of
 the Poisson bracket defined by \r{306} and
\be
&&\left.\{\cC^a(x),
\cP_b(y)\}\right|_{x^0=y^0}=-i\del^a_b\del^3(x-y),\nn\\
&&\left.\{\bcC_a(x),
\bcP^b(y)\}\right|_{x^0=y^0}=-i\del^b_a\del^3(x-y).
\e{409}
 Comparison with 
\r{312} tells us that the ghost fields $\{\cC^1,\cC^2,\bcC_1,\bcC_2\}$
 here are equal to the  ghost fields
$\{-\cC^2,-{1\over m}\bcP^2,-\cP_1,{1\over m}\bcC_1\}$ 
in the previous section. Notice that the only BRST 
invariant canonical conjugate variables are
\be
&&\tilde{A}^i\equiv A^i-{1\over m^2}\d^i\pi^0,
\quad\tilde{\pi}^i\equiv\pi^i,
\e{410}
which exactly represent the degrees of freedom of a 
massive vector field.
In the quantum theory the degrees of
 freedom are governed by BRST triplets
in involution.  As was shown in section 2 we have here two 
sets of commuting BRST triplets. They are
\be
&&1)\quad\{\bcC_1,\pi^0,m^2\cC^1\}\quad {\rm and}\quad
\{\cP_2,m\pi,m^2\bcP^2\}\nn\\
&&2)\quad\{\cP_1,\d_i\pi^i-m^2A^0,m^2\bcP^1\}\quad {\rm and}\quad
\{\bcC_2,m\varphi,m^2\cC^2\}
\e{4101} 
 or combinations of these two sets. 
They remove the $\varphi$ and $A^0$-fields and all the ghost
 variables leaving only the degrees of freedom of a
 massive vector field.

It is interesting to see what happens in the massless limit.
In the limit $m\,\ra\, 0$  
the Lagrangian \r{401} reduces to
\be
 &&\cL_0 = - {1 \over 4}F_{\mu\nu}F^{\mu\nu} -\half(\d_\mu A^\mu)^2+
i\d_\mu\bcC_1\d^\mu\cC^1+
i\d_\mu\bcC_2\d^\mu\cC^2-\nn\\
&&-\half\d_\mu \varphi\d^\mu \varphi-{1
\over 2}\mu^2 \varphi^2+i\mu^2\bcC_2\cC^2,
\e{411}
which is invariant under  the nilpotent BRST transformations
\be
&&s A^\mu=-\d^\mu\cC^1,\quad s\bcC_1=-i\d_\mu A^\mu,\nn\\
&&s\varphi=s\bcC_2=s\cC^a=0.
\e{412}
The ghosts $\cC^1$ and $\bcC_1$  are obviously reduced to the 
standard ghost, antighost fields in 
QED while $\varphi$, $\cC^2$ and $\bcC_2$ 
are turned into decoupled BRST invariant 
field variables with mass $\mu$ (BRST singlets).
The BRST current \r{405} and the BRST charge
 \r{408} reduce to the one for a massless vector field.
 Now the limit seems to lead to an inconsistent theory 
unless we remove the ghosts $\cC^2$ and $\bcC_2$.
However, in the quantum theory this is automatic since 
the two sets of commuting BRST triplets reduce
 to the BRST doublets of the massless theory 
{\em and} to BRST singlets which remove the ghosts $\cC^2$ and $\bcC_2$. 
Thus, the massless limit is consistent 
and has exactly as many degrees of freedom as
 the massive case since the scalar field $\varphi$ 
then becomes physical.

The above model is quite similar to the St\"uckelberg model
 for a free massive vector field. The latter is described by 
the Lagrangian 
\be
  &&\cL_1 = - {1 \over 4}F_{\mu\nu}F^{\mu\nu} - 
{1 \over 2}m^2 \left(A_{\mu}-{1\over m}\d_\mu\varphi\right)
\left(A^{\mu}-{1\over m}\d^\mu\varphi\right),
\e{413}
which obviously is invariant under the gauge transformation 
$A^\mu\,\ra\,A^\mu+\d^\mu\Lambda$, $\varphi\,\ra\,\varphi+m\Lambda$. 
This is an example of a conversion of second class constraints 
to first class ones. In fact, it is a particular example of 
the general conversion mechanism given in \cite{BF}. To see 
this one may consider the  equivalent Lagrangian
\be
 &&\cL_2 = - {1 \over 4}F_{\mu\nu}F^{\mu\nu} - 
{1 \over 2}m^2 A_{\mu}A^{\mu}-{1 \over 2}{\d^\mu\varphi}{\d_\mu\varphi}
-m\d_\mu A^\mu\varphi.
\e{414}
It yields the constraints
\be
&&\pi^0+m\varphi=0,\quad\d_i\pi^i-m^2A^0-m\pi=0,
\e{415}
which are the original constraints for the massive vector field obtained from
\r{300} modified  with linear terms in 
the canonical conjugate variables of the scalar
field in such a way  that the resulting constraints become  first class ones. 
The corresponding effective theory is described by the Lagrangian
\be
  &\cL_S =& - {1 \over 4}F_{\mu\nu}F^{\mu\nu} - 
{1 \over 2}m^2 A_{\mu} A^{\mu}-\half(\d_\mu A^\mu)^2-
\nn\\&&-\half\d_\mu \varphi\d^\mu \varphi-{1
\over 2}m^2 \varphi^2+i\d_\mu\bcC\d^\mu\cC+im^2\bcC\cC,
\e{416}
which is invariant under the nilpotent BRST transformation
\be
&&s A^\mu=-\d^\mu\cC,\quad s\bcC=-i(\d_\mu A^\mu-m\varphi),\nn\\
&&s\varphi=-m\cC,\quad s\cC=0.
\e{417}
The corresponding conserved BRST current is
\be
&&J^\mu=F^{\mu\nu}\d_\nu\cC-
m^2A^\mu\cC+(\d_\nu A^\nu)\d^\mu\cC+
m\varphi\stackrel{\lra}{\d^\mu}\cC.
\e{418}
This is obviously very similar properties to what we had before. 
The differences are that the number of ghost fields in 
\r{401} are twice as many as in \r{416}, the masses of the scalar
 and vector fields are the same in 
\r{416} while they may be different in \r{401}.
The St\"uckelberg model is obviously preferable at the free level
 due to its exact BRST properties. However, both models are consistent.

\setcounter{equation}{0}
\section{Conclusions}
 In the present paper we have developed the generalized BRST method for second
class constraints which was proposed in \cite{RMa,Bigrad}. This we have done by a
careful analysis of a simple model with finite number
of degrees of freedom. In this process we have
discovered what we think are rather general features of this approach.
 We have \eg demonstrated that the method works provided the number of 
unphysical bosonic
 degrees of freedom is even. (Half of the degrees of freedom must be
 quantized with indefinite metric states and half with positive 
metric states.) The physical states have the form
\be
&&|phys\hb=e^\Lambda|\phi\hb,
\e{501}
where $\Lambda$ is BRST invariant and the properties of $|\phi\hb$ is
 determined by hermitian BRST triplets ($|\phi\hb$ is then 
BRST invariant). Also in ordinary BRST quantization the physical 
states have the form \r{501} but then $|\phi\hb$ is governed by 
hermitian BRST doublets and $\Lambda$ is given by $[\psi,Q]$ 
where $\psi$ is a hermitian gauge fixing fermion. In the 
second class case there is no gauge freedom in $\Lambda$. 
However, there are different forms of $\Lambda$ which 
yields different effective theories. The precise form 
of this freedom we have not pinned down but it is 
quite limited. The 
possible forms of $\Lambda$ are connected to the precise
choice of $|\phi\hb$. In the ordinary case $|\phi\hb$ may 
be defined to be an inner product state due to the 
supersymmetry of the unphysical degrees of freedom. 
This is not the case here since the number of ghosts 
is doubled. In the considered bosonic models $|\phi\hb$ is a zero norm state.
The extra ghosts we have in the effective actions provide  for 
the necessary measure factor. The only case in which $|\phi\hb$ could be
 defined as an inner product state is when we have a supersymmetric 
set of second class constraints. Our derivations of effective 
Hamiltonians were rather awkward and cumbersome. One reason is 
that we had to use a method which is not fully developed even 
in the ordinary case. It is clear that this method may be 
considerably improved. We need simple algorithms.
 Therefore, from a mathematical point of
view there are much to be done. However, we do not expect 
that the treatment of second
class constraints can ever be as beautiful as the one for
 first class constraints,
\ie general gauge theories. 

When we applied the formalism as developed in section 2 to a free
 massive vector field we found a simple effective Lagrangian which
 was very similar to the one from the St\"uckelberg model. Notice,
 however, that the additional scalar field was in our case 
introduced in order to have inner product states while in 
the St\"uckelberg model it is introduced in order to restore 
gauge invariance. Although the St\"uckelberg formalism is
 preferable to our formulation in the free case, this might
 not be so in the interaction case. Notice that both 
formulations are consistent. Since the basic motivation behind our work was to find 
new ways to quantize massive
Yang-Mills in a consistent fashion we have at least made one step forward. 
One may notice that the generalization of the St\"uckelberg model to the 
massive Yang-Mills leads to nonpolynomial terms in the scalars which are 
nonrenormalizable \cite{ABB}. Since our model is not based on the St\"uckelberg
 transformation it is not unlikely that one may construct a massive Yang-Mills
 with polynomial interaction terms which are renormalizable. Maybe the old 
Curci-Ferrari model treated in \cite{Oj} is such a theory even when additional 
scalars are added.\\ \\

\noindent
{\bf Acknowledgements}\\

R.M. would like to thank Igor Batalin and Simon Lyakhovich
 for stimulating discussions on second
class constraints. He would also like to thank the Invitation Fellowship Program of the
Japanese Society for the Promotion of Science (JSPS) for his stay
in Japan.\\
\\

\end{document}